# The electrification of wind-blown sand on Mars and its implications for atmospheric chemistry

**Jasper F. Kok[1,2,*] and Nilton O. Renno[1,2]**
[1]Applied Physics Program, University of Michigan, Ann Arbor, MI, 48109, USA.
[2]Atmospheric, Oceanic, and Space Sciences, University of Michigan, Ann Arbor, MI, USA.

**ABSTRACT**

Wind-blown sand, or 'saltation,' creates sand dunes, erodes geological features, and could be a significant source of dust aerosols on Mars. Moreover, the electrification of sand and dust in saltation, dust storms, and dust devils could produce electric discharges and affect atmospheric chemistry. We present the first calculations of electric fields in martian saltation, using a numerical model of saltation that includes sand electrification, plasma physics, and the adsorption of ions and electrons onto particulates. Our results indicate that electric discharges do not occur in martian saltation. Moreover, we find that the production of hydrogen peroxide and the dissociation of methane by electric fields are less significant than previously thought. Both these species are highly relevant to studies of past and present life on Mars.

## 1. Introduction

Wind-blown sand, or 'saltation,' creates sand dunes and ripples, and causes wind erosion [*Bagnold*, 1941]. Moreover, the impact of saltating sand particles on the ground ejects dust aerosols into the atmosphere on both Earth and Mars [*Shao*, 2000; *Almeida et al.*, 2008], which greatly affects the climate of both planets [*Shao*, 2000; *Renno and Kok*, 2008].

Wind-blown sand and dust storms on Earth produce electric fields ($E$-fields) ranging from 1 to 200 kV/m [*Schmidt et al.*, 1998; *Renno and Kok*, 2008]. These large $E$-fields are produced by charge transfer during collisions between sand and/or dust particles and between saltating sand and the surface [*Renno and Kok*, 2008]. The physical mechanism that governs this charge transfer is poorly understood, but laboratory and field experiments indicate that saltating sand particles charge negatively while the soil surface charges positively [*Schmidt et al.*, 1998; *Kok and Renno*, 2008; *Renno and Kok*, 2008].

On Mars, electrification of wind-blown sand and dust storms could trigger electric discharges [*Eden and Vonnegut*, 1973; *Melnik and Parrot*, 1998] and reduce the wind stress required to lift particles from the surface [*Kok and Renno*, 2006]. Moreover, recent studies suggest that large $E$-fields predicted in wind-blown sand and dust storms [*Melnik and Parrot*, 1998; *Farrell et al.*, 2003; *Zhai et al.*, 2006] produce energetic electrons [*Delory et al.*, 2006] that catalyze the production of hydrogen peroxide [*Atreya et al.*, 2006], a strong oxidant hostile to life as we know it. Indeed, these studies suggest that the atmosphere becomes supersaturated, causing hydrogen peroxide snow to precipitate onto the surface [*Atreya et al.*, 2006], which provides a possible explanation for the reactive soil and the unexpected absence of organics at the Viking landing sites [*Oyama et al.*, 1977]. In addition, energetic electrons produced by strong $E$-fields are predicted to dissociate methane [*Farrell et al.*, 2006]. This is important because methane has been detected on Mars and is a possible marker of biological activity [*Formisano et al.*, 2004]. Both the production of hydrogen peroxide and the destruction of methane in martian wind-blown sand and dust storms are thus highly relevant to studies of past and present life on Mars.

In the absence of direct measurements, most researchers have used laboratory experiments and numerical models to investigate the generation of $E$-fields in martian saltation and dust storms. *Eden and Vonnegut* [1973] reported that shaking a flask of sand with $CO_2$ at martian pressure produces electric discharges. The occurrence of electric discharges in martian dust storms is also predicted by numerical studies [*Melnik and Parrot*, 1998; *Farrell et al.*, 2003; *Zhai et al.*, 2006]. However, these numerical studies have two important shortcomings. First, because the charge transfer between colliding sand/dust particles is poorly understood, these numerical studies have used charging models that are not constrained by either theory or experiments [*Renno and Kok*, 2008]. Second, these studies have neglected the effects of $E$-fields on atmospheric conductivity. Fortunately, progress has recently been made on both these issues. Indeed, *Delory et al.* [2006] developed a plasma physics model that accounts for the production of energetic electrons by $E$-fields and the subsequent ionization of martian air, while we recently developed an improved parameterization of sand/dust electrification that is constrained by $E$-field measurements in saltation on Earth [*Kok and Renno*, 2008].

In this Letter, we build on the studies of *Delory et al.* [2006] and *Kok and Renno* [2008] and report the first calculations of $E$-fields in martian saltation. Our study is an improvement over calculations of *E*-fields in dust storms [*Melnik and Parrot*, 1998; *Farrell et al.*, 2003; *Zhai et al.*, 2006] for three reasons: (i) our parameterization of the charge transfer between colliding sand/dust particles is constrained by measurements [*Kok and Renno*, 2008], (ii) we account for the effects of *E*-fields on atmospheric conductivity [*Delory et al.*, 2006], and (iii) we account for the adsorption of ions and electrons to particulates [*Draine and Sutin*, 1987; *Jackson et al.*, 2008]. We find that electric discharges are unlikely to occur in martian wind-blown sand, and that the production of hydrogen peroxide and the dissociation of methane by *E*-fields are less significant than previously thought.

**2. Model description**

Our numerical model of saltation is described in detail in *Kok and Renno* [2008], and explicitly simulates the motion, concentration, and electric charging of saltating sand. Our model simulates saltation in the absence of suspended dust, as is representative of saltation on dunes, and its predictions are in good agreement with measurements of the particle mass flux and *E*-field in terrestrial saltation [*Kok and Renno*, 2008]. Here, we apply our model to Mars and calculate the *E*-field in saltation as described in *Kok and Renno* [2008]. We assume that saltating particles have diameters $D_p$ = 100 μm and density of 3000 kg/m$^3$ [*Claudin and Andreotti*, 2006], and take the atmospheric pressure (*P*) and temperature (*T*) as 627 Pa and 227 K. As described in more detail below, we expand the model by including the effects of *E*-fields on atmospheric conductivity and accounting for the adsorption of ions and electrons to particulates.

**2.1 Limits to Electric Fields on Mars**

On Earth, sand and dust electrification can produce large *E*-fields [*Renno and Kok*, 2008] because air is a good insulator and the *E*-field at which electric discharges occur is large (about 3 MV/m). The situation is quite different on Mars. There, *E*-fields are limited by large increases in atmospheric conductivity when *E*-fields become sufficiently large to ionize $CO_2$ [*Delory et al.*, 2006], and by electric discharges thought to occur at ~20-25 kV/m [*Melnik and Parrot*, 1998].

The *E*-field at which the insulating properties of a gas break down and an electric discharge occurs is described by the 'Paschen law' [*Raizer*, 1997; *Fridman and Kennedy*, 2004], and depends on the gas pressure and the distance of the "electrodes" (or centers of charge) between which the discharge occurs,

$$E_{br} = \frac{BPT_0/T}{C + \ln(Pz_{cat}T_0/T)}, \quad (1)$$

with $C = \ln[A/\ln(1/\gamma + 1)]$. The constants $A$ = 15 m$^{-1}$Pa$^{-1}$ and B = 350 Vm$^{-1}$Pa$^{-1}$ define the Townsend ionization coefficient α (see page 56 in *Raizer* [1997]) at $T_0$ = 293 K for a $CO_2$ atmosphere. We take the secondary Townsend ionization coefficient γ as 0.01 [*Raizer*, 1997; *Fridman and Kennedy*, 2004]. Note that equation (1) does not include the effect of sand and dust on the breakdown *E*-field. In the case of negatively charged saltating sand over a positively charged soil surface [*Kok and Renno*, 2008], the surface represents the anode, but the cathode is not well defined. We approximate the distance from the cathode to the anode by the height $z_{cat}$ below which 50 % of the charge on saltating sand is contained. The results reported here are not sensitive to this approximation. For typical martian saltation, we find $z_{cat}$ = 30 cm and $E_{br}$ = 43 kV/m, which is significantly above the ~20-25 kV/m value at which larger-scale discharges in dust storms are thought to occur [*Melnik and Parrot*, 1998].

The second mechanism limiting the generation of *E*-fields in martian saltation and dust storms is the increase in atmospheric conductivity due to ionization by energetic electrons [*Delory et al.*, 2006]. The conductivity of the near-surface martian atmosphere is due mostly to mobile ions [*Molina-Cuberos et al.*, 2002], and equals [e.g., *Michael et al.*, 2008]

$$\sigma = e(K_e n_e + K_- n_- + K_+ n_+),  \qquad (2)$$

where *e* is the elementary charge, and $K_e$, $K_-$, $K_+$, $n_e$, $n_-$, and $n_+$ are the mobilities and number densities of free electrons and negative and positive ions, respectively. We take the 'background' concentrations of electrons and ions to be $n_{e,0} = 5 \times 10^6$ m$^{-3}$ and $n_{-,0} = n_{+,0} = 3 \times 10^9$ m$^{-3}$ [*Molina-Cuberos*, 2001, 2002; *Delory et al.*, 2006]. Charges in the martian atmosphere decay due to the adsorption of electrons and ions of opposite polarity, which is a complex process [*Draine and Sutin*, 1987; *Michael et al.*, 2008]. However, the conductivity defines the approximate time scale $t_{rel} = \varepsilon_0 / \sigma$, where $\varepsilon_0$ is the electric permittivity, with which charges in the martian atmosphere decay. A simplified expression of this charge decay is thus

$$q(t + \Delta t) = q(t) \exp(-\Delta t / t_{rel}),  \qquad (3)$$

where *q*(*t*) is the charge of the particle (or the surface) at time *t*, and Δ*t* is the model time step. As the atmospheric conductivity increases, the charge relaxation time $t_{rel}$ decreases, thereby also decreasing the charge held by saltating particles and the soil surface.

**2.2 Plasma Physics**

Electric fields on Mars are thus limited by the occurrence of electric discharges (Equation 1) and by increases in atmospheric conductivity (equations 2 and 3). As electrons are accelerated from the top cathode (the top of the saltation layer) towards anode (the surface), they can ionize $CO_2$ and produce additional free electrons, but can also be absorbed through dissociative attachment to $CO_2$ [*Delory et al.*, 2006] and collisions with saltating sand particles. The electron concentration in the saltation layer is then approximately given by [*Raizer*, 1997; *Delory et al.*, 2006; *Jackson et al.*, 2008]

$$n_e = n_{e,0} \exp \left\{ \int_0^{z_{cat}} \left[ \alpha(z) - \frac{k_{da}(z) N_{CO2}}{v_d(z)} - \frac{1}{4} \pi D_p^2 n_{salt}(z) \tilde{J}(z) \right] dz \right\},  \qquad (4)$$

where $N_{CO2}$ is the $CO_2$ number concentration, $v_d$ is the electron drift velocity and is obtained from Figure 4a in *Delory et al.* [2006], $n_{salt}$ is the concentration of saltating sand particles as predicted by our saltation model [*Kok and Renno*, 2008], and $\tilde{J}$ is the normalized cross section for a collision occurring between an electron and a saltating sand particle. Since the sand particles are strongly negatively charged, we have $\tilde{J} < 1$, following equation (3.5) in *Draine and Sutin* [1987] and using equation (1) in *Jackson et al.* [2008] to obtain the electron temperature as a function of the *E*-field. Furthermore, we obtain the dissociative attachment rate constant $k_{da}$ from Figure 4d of *Delory et al.* [2006], who solve the electron energy distribution to find $k_{da}$ as a function of the *E*-field. Finally, we use Figure 4b of *Delory et al.* [2006] to obtain the Townsend ionization coefficient *α*, which describes the multiplication of electrons per unit length due to ionizing collisions as the initial population ($n_{e,0}$) is accelerated from the cathode to the anode [*Raizer*, 1997; *Fridman and Kennedy*, 2004]. This electron population becomes increasingly energetic as the *E*-field rises, and can produce positive ions (mainly $CO_2^+$) through electron impact ionization, and negative ions (mainly $O^-$) through dissociative attachment [*Delory et al.*, 2006]. The $CO_2^+$ ions quickly react with $CO_2$, $O_2$, and $H_2O$ to form $H_3O^+ \cdot (H_2O)_j$ with j ≥ 1, while $O^-$ ions attach to $CO_2$, forming $CO_3^-$, which is hydrated to $CO_3^- \cdot (H_2O)_j$ [*Molina-Cuberos et al.*, 2001, 2002]. The concentration of negative ions in the saltation layer is then described by [*Michael et al.*, 2008]

$$\frac{dn_-}{dt} = \frac{N_{CO2} n_e}{z_{cat}} \int_0^{z_{cat}} k_{da}(z) dz - n_- n_+ k_{rec} - \frac{n_- K_- E_{surf}}{z_{cat}}, \quad (5)$$

where $N_{CO2}$ is the $CO_2$ number concentration, and $E_{surf}$ is the surface *E*-field. The first term on the right-hand side denotes the production of negative ions through dissociative attachment to $CO_2$ [*Delory et al.*, 2006], the second term describes the recombination of positive and negative ions [*Molina-Cuberos et al.*, 2002], and the final term accounts for the adsorption of negative ions to the positively charged soil surface. We neglect other processes that are insignificant compared to these ion loss processes, such as photodetachment [*Molina-Cuberos et al.*, 2001; *Michael et al.*, 2007] and the attachment of negative ions to the strongly negatively charged saltating grains [*Draine and Sutin*, 1987], and we neglect the transport of ions out of the saltation layer. Furthermore, we take the ion recombination rate constant as $k_{rec}$ = 1.5×10$^{-13}$ m$^3$sec$^{-1}$ [*Molina-Cuberos et al.*, 2002], and calculate $n_-$ and $n_+$ iteratively using $n_+ = n_- + n_e$ [*Molina-Cuberos et al.*, 2001] and $dn_-/dt = 0$ in steady-state. We then use the ion and electron concentrations to calculate the atmospheric conductivity using (2), with ion and electron mobilities derived from equations (5) and (7) of *Michael et al.* [2008], assuming the dominant ions to be $H_3O^+ \cdot (H_2O)_3$ and $CO_3^- \cdot (H_2O)_2$ [*Molina-Cuberos et al.*, 2001, 2002].

### 3. Results and Discussion

We implement the plasma physics processes discussed above in the numerical model of saltation described in [*Kok and Renno*, 2008], and iteratively calculate the *E*-field, the atmospheric conductivity, and the motion, charging, and concentration of saltating sand until steady-state is reached. As on Earth [*Kok and Renno*, 2008], the *E*-field in martian saltation peaks at the surface and decreases monotonically with height (inset of Figure 1). On Mars, the rate of decrease of the *E*-field with height is less than on Earth because the smaller gravitational and aerodynamic drag forces cause the saltation layer to be thicker there [*Almeida et al.*, 2008].

As expected, the *E*-field in the saltation layer increases with wind speed (Figure 1). The resulting increasingly energetic electron population starts dissociating $CO_2$ (i.e., $e + CO_2 \rightarrow CO + O^-$) at a few kV/m and ionizing $CO_2$ (i.e., $e + CO_2 \rightarrow 2e + CO_2^+$) at ~10 kV/m [*Delory et al.*, 2006]. The resulting increase in the concentration of ions and electrons with the *E*-field (Figure 2) enhances the atmospheric conductivity, which neutralizes the charges on saltating particles and the surface (see equation 3), thereby limiting further increases in the *E*-field. Indeed, we find that this negative feedback limits the *E*-field in martian saltation to ~15–20 kV/m. This upper limit on the *E*-field in martian saltation is relatively insensitive to uncertainties in model parameters and the model methodology, because of the sharp dependence of the production rate of ions on the *E*-field (see *Delory et al.* [2006], figure 4d).

Since the maximum *E*-field of ~15-20 kV/m is significantly below the threshold of ~43 kV/m required to initiate electric discharges (see equation 1), we conclude that such discharges are unlikely to occur in martian saltation. However, discharges might still occur in dust devils and dust storms for several reasons. First, the *E*-field required to precipitate discharges over larger scales in dust storms is lower (see equation 1). Moreover, the abundant presence of particulate matter in dust storms likely lowers the background concentration of ions and electrons [*Eden and Vonnegut*, 1973; *Michael et al.*, 2008], thereby increasing the charge relaxation time and thus the *E*-field. Finally, large-scale discharges in dust storms could occur at a lower *E*-field than predicted by the Paschen law (equation 1) through electron runaway breakdown [*Gurevich et al.*, 1992].

Recent studies have predicted that *E*-fields of 10-25 kV/m generate plasma conditions that produce hydrogen peroxide and dissociate methane in large quantities [*Delory et al.*, 2006; *Atreya et al.*, 2006; *Farrell et al.*, 2006]. While we here indeed find that the *E*-field in saltation can exceed 10 kV/m for large wind speeds, we also find that the concentration of ions and electrons at such *E*-fields is much smaller than suggested by these previous studies (Figure 2). The difference occurs because we expanded on these previous studies and accounted for losses of ions and electrons due to adsorption to saltating sand and the soil surface, as well as the loss of electrons from dissociative attachment to $CO_2$. A separate calculation shows that the large concentration of electrons and ions predicted by following these previous studies [*Delory et al.*, 2006; *Atreya et al.*, 2006] are unlikely to occur in saltation or dust storms, because the large conductivity of the resulting plasma limits the *E*-field to values well below those necessary to maintain the plasma (Figure 3). Indeed, the charging current necessary to maintain these plasma conditions is several orders of magnitude larger than that produced by saltation (Figure 3). Since saltation probably plays a key role in charge generation in dust storms and dust devils [*Renno and Kok*, 2008], we expect the charging current in these phenomena to be of similar magnitude as in saltation. We therefore conclude that the concentration of ions and electrons in martian wind-blown sand, dust devils, and dust storms, is much smaller than previously suggested [*Delory et al.*, 2006; *Atreya et al.*, 2006]. The production of hydrogen peroxide and the dissociation of methane by *E*-fields in these phenomena are thus probably less significant than previously thought [*Atreya et al.*, 2006; *Farrell et al.*, 2006].

## 4. Conclusions

We present the first numerical simulation of *E*-fields in martian saltation, and find that *E*-fields are limited to ~15-20 kV/m. This upper limit is imposed by the rapid increase in atmospheric conductivity as the *E*-field rises and probably prevents the occurrence of electric discharges in martian saltation.

Furthermore, our results show that chemical reactions catalyzed by *E*-fields in saltation are not as important as previously thought [*Atreya et al.*, 2006; *Farrell et al.*, 2006]. Indeed, we find that the plasma in which these reactions occur cannot be sustained because its large conductivity limits the *E*-field to values well below that necessary to maintain the plasma (Figure 3). Nonetheless, the concept of electro-chemical production of oxidants in martian saltation and dust storms, possibly through electric discharges, remains a possible explanation for the puzzling absence of organics from the martian soil [*Oyama et al.*, 1977; *Atreya et al.*, 2006] and should be investigated further.

**Acknowledgements**

We thank John Barker and Robert Sullivan for useful discussions, Shanna Shaked for comments on the manuscript, and NSF and NASA for financial support through awards ATM 0622539 and NNX07AM99G.

**Figures**

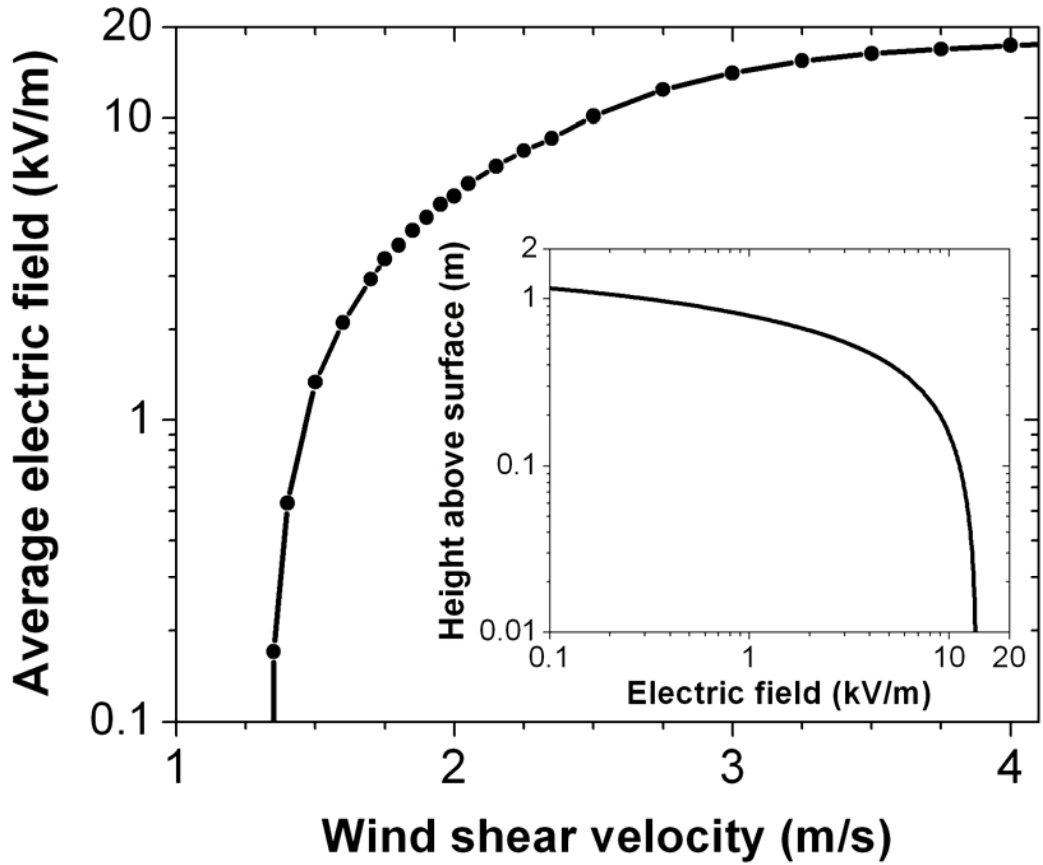

**Figure 1.** Simulated average *E*-field between the anode (the surface) and the cathode ($z_{cat}$ ≈ 30 cm) in Martian saltation as a function of wind shear velocity, $u^* = \sqrt{\tau/\rho_a}$, where $\tau$ is the wind shear stress directly above the saltation layer [*Shao*, 2000] and $\rho_a$ is atmospheric density. The inset shows the vertical profile of the *E*-field for a wind shear velocity of 2.5 m/s. The results are obtained with the numerical model of saltation described by *Kok and Renno* [2008], expanded with equations (2)-(5) to account for plasma physics processes and charge relaxation.

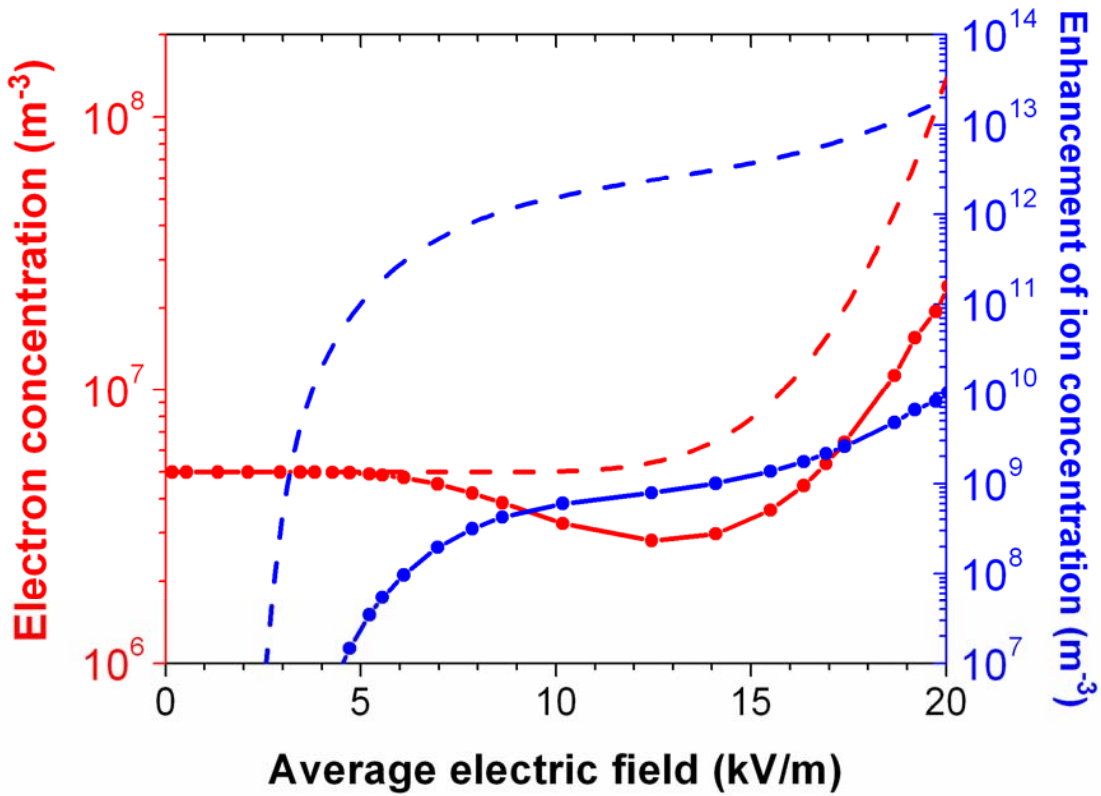

**Figure 2.** Simulated concentration of electrons (left axis and red lines) and the enhancement of the ion concentration over the background concentration ($n_{-,0} = n_{+,0} = 3 \times 10^9$ m$^{-3}$ [*Molina-Cuberos et al.*, 2002]; right axis and blue lines) as a function of the average *E*-field between the anode and the cathode. Solid lines with circles indicate results from our numerical saltation model [*Kok and Renno*, 2008], for which the anode is at the surface and the cathode is at the height $z_{cat}$ (see text). For *E*-fields of ~5-12 kV/m, the electron concentration decreases because of dissociative attachment to $CO_2$ and adsorption to sand particles, whereas for larger *E*-fields the electron concentration increases due to the generation of additional electrons through ionization of $CO_2$ [*Delory et al.*, 2006]. Dashed lines indicate electron and ion concentrations calculated for a homogenous *E*-field over a length of 0.5 m following *Delory et al.* [2006] and *Atreya et al.* [2006]. That is, we use equations (4) and (5) to calculate the ion and electron concentrations, but neglect the terms in these equations that account for losses of electrons and the loss of ions to the soil surface.

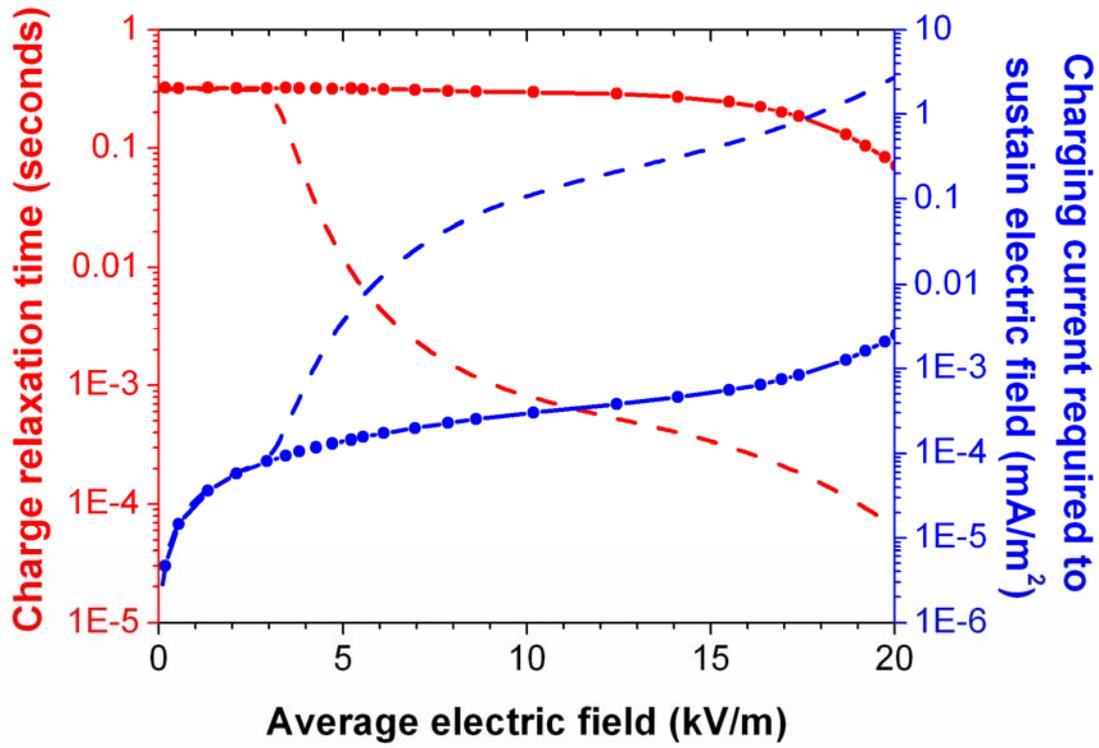

**Figure 3.** Simulated charge relaxation time (left axis and red lines) and charge separation current ($I = \sigma E$) required to sustain the $E$-field (right axis and blue line) against the relaxation of charge on saltating particles and the surface, as a function of the average $E$-field between the anode and the cathode. Solid lines with circles indicate results from our numerical saltation model [*Kok and Renno*, 2008] for which the anode is at the surface and the cathode is at the height $z_{cat}$ (see text). Dashed lines indicate results following *Delory et al.* [2006] and *Atreya et al.* [2006], as described in the caption of Figure 2.